\documentclass[nohyperref]{article}

\usepackage{microtype}
\usepackage{graphicx}
\usepackage{subfigure}
\usepackage{booktabs} 

\usepackage{hyperref}

\usepackage[accepted]{icml2022}

\usepackage{amsmath}
\usepackage{amssymb}
\usepackage{mathtools}
\usepackage{amsthm}
\usepackage{hyperref}

\usepackage[capitalize,noabbrev]{cleveref}

\DeclareMathOperator*{\argmin}{argmin}

\theoremstyle{plain}
\newtheorem{theorem}{Theorem}[section]
\newtheorem{proposition}[theorem]{Proposition}

\theoremstyle{definition}
\newtheorem{definition}[theorem]{Definition}

\theoremstyle{remark}
\newtheorem{remark}[theorem]{Remark}

\usepackage[textsize=tiny]{todonotes}

\icmltitlerunning{An Optimal Likelihood Free Method for Biological Model Selection for ICML-WCB 2022}

\begin{document}

\twocolumn[
\icmltitle{An Optimal Likelihood Free Method for Biological Model Selection}



\begin{icmlauthorlist}
\icmlauthor{Vincent D. Zaballa}{yyy}
\icmlauthor{Elliot E. Hui}{yyy}
\end{icmlauthorlist}

\icmlaffiliation{yyy}{Department of Biomedical Engineering, University of California Irvine, Irvine, CA, USA}

\icmlcorrespondingauthor{Vincent Zaballa}{vzaballa@uci.edu}
\icmlcorrespondingauthor{Elliot Hui}{eehui@uci.edu}

\icmlkeywords{Biology, Bayesian, Machine Learning, ICML}

\vskip 0.3in
]



\printAffiliationsAndNotice{}  

\begin{abstract}
Systems biology seeks to create math models of biological systems to reduce inherent biological complexity and provide predictions for applications such as therapeutic development. However, it remains a challenge to determine which math model is correct and how to arrive optimally at the answer. We present an algorithm for automated biological model selection using mathematical models of systems biology and likelihood free inference methods. Our algorithm shows improved performance in arriving at correct models without a priori information over conventional heuristics used in experimental biology and random search. This method shows promise to accelerate biological basic science and drug discovery.
\end{abstract}

\section{Introduction}
\label{submission}

Biological cellular systems exhibit super exponential scaling in the number of biological states achieved arising from different combinations and sequences of cell regulators, such as messenger proteins and transcription factors \cite{Letsou2016}. This complexity impedes our understanding of diseases and development of therapeutics. We focus here on the combinatorial complexity of biology, defined by the vast number of models and their parameters that describe biological systems. 

This combinatorial problem in biology is exemplified by promiscuous signaling, which is the phenomenon of multiple protein ligands in a pathway being able to bind to multiple receptors in a competitive manner. The Bone Morphogenetic Protein (BMP) pathway exemplifies this type of signaling with multiple protein ligands, and type I and II receptors present in the pathway, each combining with one another at different rates to form a complex of ligand, type I, and type II receptor to phosphorylate SMAD 1/5/8 to send a downstream gene expression signal. The BMP pathway can be mathematically modeled by mass action kinetics \cite{Antebi2017} and previous work demonstrated how to optimally infer BMP models' parameters using Likelihood Free Inference (LFI), also known as Simulation Based Inference (SBI), using the SBIDOEMAN algorithm \cite{Zaballa2021}. However, since multiple models have been proposed for the BMP pathway \cite{Antebi2017, Su2022}, there remains ambiguity in determining which model best describes observed experimental data. 

We propose to use the previously developed SBIDOEMAN algorithm and a novel method to approximate a model's marginal probability, $p(\mathcal{M} | \boldsymbol{x_o}, \boldsymbol{\theta})$, within Bayesian Model Averaging (BMA) to select a correct model from a set of models proposed. This method, which we call SBIDOEMAN BMA, uses models' prior distributions of parameters, $p(\boldsymbol{\theta})$, to design optimal experiments using a mutual information approximation $I(\boldsymbol{\theta}, \boldsymbol{x}; d)$ between model parameters and data, then determines the posterior distribution of parameters given observed data, $p(\boldsymbol{\theta}|\boldsymbol{x_o})$, by LFI, and finally approximates a marginal likelihood of a biological model given observed data points, $p(\mathcal{M} | \boldsymbol{x_o}, \boldsymbol{\theta})$. This marginal probability is used as a probability measure of a given model, $\mathcal{M}$, and can be used in BMA to determine the next experiment to evaluate and a weighting of possible models. 

Previous work for optimal experimental designs in biological systems studied graphical models describing gene regulatory networks, modeled using Bayesian graphs, \cite{Cho2016} and M-estimators applied to Gaussian Markov Random fields, \cite{Zheng2018} both of which have closed-form information measures. By contrast, we evaluate methods exclusive to the LFI setting where likelihoods and closed-form information measures are not tractable. Regarding model selection, trained classifiers have been proposed to classify whether data can fit a proposed model or not \cite{Radev2021}. While useful in model selection, this method does not provide a posterior distribution of models' parameters or design optimal experiments.  Our method provides an alternative for scientists who have a model of their system that they can simulate but not evaluate its likelihood function, compare models, and design experiments towards the most promising model. Additionally, our method has the potential to be used with biological highthroughput screening (HTS) systems to increase the efficiency of such systems to discover novel biology and therapeutics. 

In summary, the key contributions of this paper are:

\begin{itemize}
    \item A method to determine the marginal probability of a model given observed data.
    \item BMA applied to optimal experimental designs to design experiments for a given model.
    \item An automated algorithm to design and evaluate experiments in biological models that is compatible with HTS of biological systems.
\end{itemize}

\section{Background}

\subsection{Modeling the BMP pathway}

Two mass action kinetics models have been proposed for the BMP pathway. The one-step model in \cref{one-step} models type I (A) and type II (B) receptors and a ligand (L) forming a trimer complex in a single step \cite{Su2022}
\begin{align}\label{one-step}
    A + B + L \xrightarrow{K} T.
\end{align}

The two-step model in \Cref{2,3} adds a parameter to model a ligand first binding with a type I receptor before forming a trimeric complex with a type II receptor \cite{Antebi2017} as follows
\begin{align}
    \label{2}A + L \xrightarrow{K_1} D \\
    \label{3}B + D \xrightarrow{K_2} T.
\end{align}

Both models have a complex, $T$, that phosphyrolyates SMAD to send a downstream gene expression signal, $S$, with a certain efficiency, $\epsilon$ as
\begin{align}\label{PhosEfficiency}
    \epsilon T = S.
\end{align}

Steady-state signals can be simulated using convex optimization \cite{Su2022}.

\subsection{Normalizing Flows}

Given a dataset, one may ask what is the probability of a certain data point in the dataset, $p_x(\boldsymbol{x})$, of a variable $\boldsymbol{x}$ with $\mathbb{R}^D$ dimensions. However, this probability density is usually intractable or unknown. Normalizing flows provide a way to answer this question by creating a transformation from a known simple distribution, $p_u(\boldsymbol{u})$, such as a Gaussian distribution, to the data distribution, $p_x(\boldsymbol{x})$, by a series of nonlinear and invertible composition of functions, $\boldsymbol{f} : \mathbb{R}^D \rightarrow \mathbb{R}^D$, where $\boldsymbol{f}$ is composed of $N$ functions, $\boldsymbol{f} = \boldsymbol{f}_N \circ \dots \circ \boldsymbol{f}_1$. We can map from a base distribution to target distribution using the change-of-variables formula for random variables as 
\begin{align}\label{NormFlow}
    p_x(\boldsymbol{x})  = p_u(\boldsymbol{u}) \rvert \det \boldsymbol{J}(\boldsymbol{f})(\boldsymbol{u}) \rvert ^{-1},
\end{align}

where $\boldsymbol{J}(\boldsymbol{f})(\boldsymbol{u})$ is the Jacobian matrix of $\boldsymbol{f}$ evaluated at $\boldsymbol{u}$. See \citet{pml2Book} for details about normalizing flows.

\subsection{Likelihood Free Inference}

For models with an implicit or intractable likelihood function, $p(\boldsymbol{x}|\boldsymbol{\theta})$, but whose response may be simulated, we can use LFI methods to approximate the posterior $q(\boldsymbol{\theta}|\boldsymbol{x})$ or likelihood $q(\boldsymbol{x}|\boldsymbol{\theta})$. This can be done by drawing $N$ samples from the prior $p(\boldsymbol{\theta})$ and generating a dataset $\{(\boldsymbol{\theta}_n, \boldsymbol{x}_n)\}^N_{n=1}$ by sampling $\boldsymbol{\theta}_n \sim p(\boldsymbol{\theta})$. Each $(\boldsymbol{\theta}_n, \boldsymbol{x}_n)$ is a joint sample from $p(\boldsymbol{\theta}, \boldsymbol{x}) = p(\boldsymbol{\theta})p(\boldsymbol{x}|\boldsymbol{\theta})$, and can be used to train a normalizing flow to approximate the posterior $q(\boldsymbol{\theta}|\boldsymbol{x})$ conditioned on an observed $\boldsymbol{x_o}$ \cite{Greenberg2019, Papamakarios2016} or approximate the likelihood $q(\boldsymbol{x}|\boldsymbol{\theta})$ conditioned on $\boldsymbol{\theta}$. See \citet{Papamakarios2019} for details on applying normalizing flows to LFI. 

While LFI provides a method to approximate a model's posterior or likelihood, practical considerations, such as difficulty in rejection sampling in in sequential neural posterior estimate (SNPE) \cite{Greenberg2019} or prohibitively slow MCMC sampling for sequential neural likelihood estimate (SNLE) \cite{Papamakarios2018}, make LFI methods difficult to implement. In response to this difficulty, recent methods have developed variational methods to approximate the posterior or likelihood \cite{Glockler2022, Wiqvist2021}. These methods, referred to here as sequential neural likelihood variational inference (SNLVI), train another normalizing flow, $q_{\phi}(\boldsymbol{\theta})$, to minimize the divergence from an estimated likelihood, $\phi^* = \argmin_{\phi} D(q_{\phi}(\boldsymbol{\theta}) ||  q_{\psi}(\boldsymbol{x}|\boldsymbol{\theta}))$. We use SNLVI methods to overcome prior practical difficulties in LFI methods.

\subsection{Optimal Experimental Design for Implicit Likelihood Model Selection}

Optimal experimental designs (OEDs) can be formulated as an optimization \cite{Boyd} or information theoretic problem \cite{MacKay1992}. Assuming designs are independent of model parameters, we formulate this problem as maximizing the information gain (IG) \cite{Foster2019a}, or, the difference in entropy given a proposed design, $d$, as
\begin{align}\label{DOE}
    \mathrm{IG}(\boldsymbol{x}, \boldsymbol{d}) = H[p(\boldsymbol{\theta})] - H[p(\boldsymbol{\theta}|\boldsymbol{x}, \boldsymbol{d})].
\end{align}

This objective function can be rewritten as a utility function, $U(\boldsymbol{d})$, that maximizes the mutual information (MI), $I(\boldsymbol{v}; \boldsymbol{y}|\boldsymbol{d})$ between a variable of interest, $\boldsymbol{v}$, and the observed data, $\boldsymbol{x}$, at particular design, $\boldsymbol{d}$. The MI variable of interest, $\boldsymbol{v}$, can be adapted to the scientific question at hand \cite{Ryan2016}. A gradient-based approach for OEDs was recently proposed for likelihood free models that provides a way to both select a model, $\mathcal{M}$, by BMA and determine its parameters, $p(\boldsymbol{\theta} | \mathcal{M})$ with a minimum number of experiments \cite{Kleinegesse2021}. Finding designs that optimally discover a model and its parameters can be formulated as the following utility function
\begin{equation}
\begin{split}\label{MDPE}
    U(\boldsymbol{d}) = \sum_{\mathcal{M}} \int p(\mathbf{x}|\boldsymbol{\theta}_{\mathcal{M}}, \mathcal{M}, \mathbf{d}) p(\boldsymbol{\theta}_{\mathcal{M}}, \mathcal{M}) \\ \log \left( \frac{p(\boldsymbol{\theta}_{\mathcal{M}}, \mathcal{M}| \mathbf{x}, \mathbf{d})}{p(\boldsymbol{\theta}_{\mathcal{M}}, \mathcal{M})} \right) d\mathbf{x}.
\end{split}
\end{equation}

We implement equation 7 by simply averaging each model's Mutual Information Neural Estimation (MINE) \cite{Belghazi2018} MI estimate. The estimated MI is then used as the objective function in Bayesian Optimization using a Gaussian Process \cite{Kleinegesse2020b}.

\subsection{Bayesian Model Averaging and the Bayes Factor}

The weighting of model probabilities is also known as the Bayes Factor (BF), which we define here as $BF = p(\mathcal{M}_1) / p(\mathcal{M}_0)$, and can be used as a form of model selection where $BF > 10$ is strong evidence for $\mathcal{M}_1$ and $BF < 1/10$ is strong evidence for $\mathcal{M}_0$. We only use the BF for model selection as it uses marginal probabilities that prefer simpler models by the Bayesian Occam's razor effect. Although, this relies on an accurate estimate of the model's marginal probability. See \citet{pml1Book} for further discussion on various model selection techniques.

\subsection{Approximating Model Marginal Probability}

To perform model selection, we need an estimate of each model's marginal probability in order to calculate the BF. To do this, we can use a normalizing flow with a Gaussian base distribution $p_u(\boldsymbol{u})$ that can provide a probability of a model given the posterior parameter distribution and observed data, $p(\mathcal{M} | \boldsymbol{x_o}, \boldsymbol{\theta}, \boldsymbol{d})$, which is the same as marginal likelihood, $p(\boldsymbol{x_o}|\boldsymbol{\theta}, \mathcal{M}, \boldsymbol{d})$, when assuming uniform priors over models, $p(\mathcal{M}_i) = 1/ |\mathcal{M}|$. This flow is trained by sampling data from the simulator of $\mathcal{M}$ to produce $\boldsymbol{x} \sim p_x(\boldsymbol{x}|\boldsymbol{x_o}, \mathcal{M}, \boldsymbol{\theta})$ that can be used to train a reverse flow function to a base Gaussian distribution $\boldsymbol{u} = \boldsymbol{f}^{-1}(\boldsymbol{x})$. We propose the following method to approximate the marginal likelihood. 
\begin{proposition}
\label{prop:margProbFlow}
The marginal likelihood of a model, $\mathcal{M}$, given an observed data vector, $\boldsymbol{x_o}$, and the model's parameters, $\boldsymbol{\theta}$, can be approximated as $p(\boldsymbol{x_o}|\mathcal{M}) \approx 1 - \Phi(\boldsymbol{f}^{-1}(\boldsymbol{x_o}))$, where $\boldsymbol{f}^{-1}$ is the pullback of a trained normalizing flow from the observed data distribution, $p_x(\boldsymbol{x_o})$, to a Gaussian base distribution, $p_u(\boldsymbol{u})$, and $\Phi$ is cumulative distribution function of a Gaussian distribution.  
\end{proposition}

We provide a proof of \cref{prop:margProbFlow} in Appendix B. \cref{sdmAlgo} in Appendix A brings these parts together in SBIDOEMAN BMA to optimally determine models and their parameters.
\begin{table}[ht]
\begin{center}
\begin{small}
\begin{sc}
\begin{tabular}{lcccr}
\toprule
Policy & Median BF & 25\% & 75\%  \\
\midrule
one-step Random   & 0.05 & 0.02 & 0.17 \\
one-step Equi     & 0.55 & 0.09 & 3.72\\
\textbf{one-step SDM BMA}      & $\bold{0.03}$ & $\bold{0.01}$ & $\bold{0.05}$ \\
\hline
two-step Random   & 0.74 & 0.22 & 1.28      \\
two-step Equi     & 2.12 & 0.79 & 16.11\\
\textbf{two-step SDM BMA}      & $\bold{5.70}$ & $\bold{1.38}$ & $\bold{34.66}$ \\
\bottomrule
\end{tabular}
\end{sc}
\end{small}
\end{center}
\caption{Median and interquartile range (IQR) Bayes Factor (BF) values after 5 rounds of experiments for both one-step and two-step datasets compared to random and equidistant experimental design policies. Lower BF is better for the series of one-step models while higher BF is better for the two-step model. For both models, both the median and IQR values are better than competing approaches.}
\vskip -0.6pt
\label{IQR-table}
\end{table}
\begin{figure}[ht]
\begin{center}
\centerline{\includegraphics[width=\columnwidth]{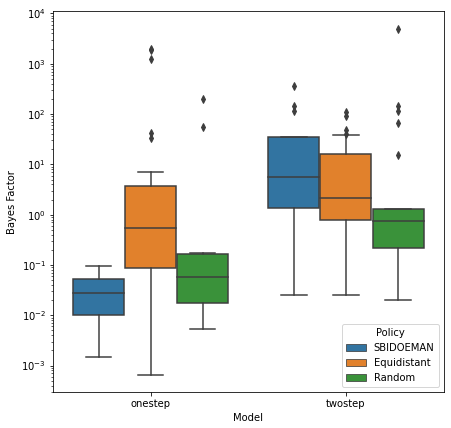}}
\caption{Final Bayes Factor (BF) after 5 design rounds and an ensemble of models. Compared to controls for both models, SBIDOEMAN BMA performed an order of magnitude better on the one-step model and performed more than two times better than control policies of the two-step model.}
\label{boxplot-comparison}
\end{center}
\vskip -0.4in
\end{figure}

\begin{figure}[ht]
\centerline{\includegraphics[width=\columnwidth]{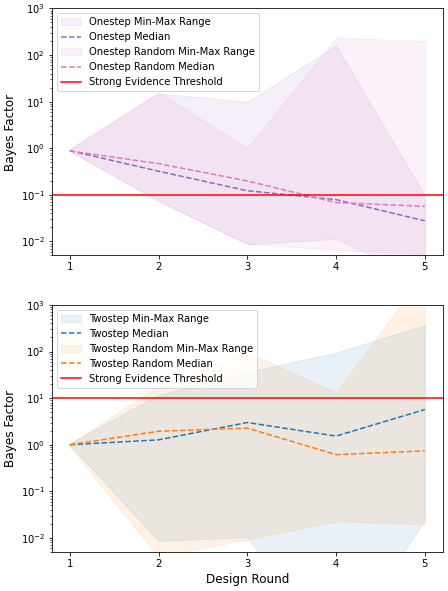}}
\caption{Change in Bayes Factor (BF), $p(\mathrm{twostep})/p(\mathrm{onestep})$, over design round when the one-step (top) and two-step (bottom) models are true. The strong evidence threshold for both models is labeled in red.
\textcolor{blue}{Top}: When the one-step model is true, SBIDOEMAN BF model trends down, indicating the one-step model is true and outperforms random search by the final design. The median BF value for the SBIDOEMAN model strongly suggests the one-step model is true by the fifth round.
\textcolor{red}{Bottom}: When the two-step model is true the median value of the SBIDOEMAN BF trends upwards, indicating the two-step model is true, and has a median trend that outperforms the competing random search by the last three designs. The two-step model's final value indicates only moderate evidence in favor of the true two-step model. }
\label{BF-trend}
\vskip -0.2in
\end{figure}

\section{Results}
We evaluated SBIDOEMAN BMA for model selection by evaluating the BF over five rounds of experiments when the one-step BMP pathway was true and when the two-step BMP pathway was true by holding out a single set of parameters for each model, $\boldsymbol{\theta}_T^{\{1,2\}}$. When evaluating performance across designs, we compared to random search, as shown in \cref{BF-trend}. We also compared final BF to random and equidistant ligand titrations which is a heuristic commonly used in biology to evaluate the response of an assay. Equidistant designs are logarithmically-equal spaced designs across a domain of interest. Here, this would be five equally spaced designs in concentrations from $10^{-3}$ to $10^{3} \mathrm{ng/ml}$. Results of the final design comparison are shown in \cref{boxplot-comparison} and \cref{IQR-table}. 

Examining the change in BF across designs \cref{BF-trend}, we see that across an ensemble of independent and identically distributed (iid) SBIDOEMAN models that the median performance outperforms random search for both the one-step and two-step models. When looking at the final BF after a budget of 5 designs, as shown in Table 1 and Figure 2, we see that the median performance of SBIDOEMAN BMA outperforms random and equidistant data, with SBIDOEMAN BMA interquartlie range (IQR) values performing better, or almost better, than competing policy median values. While random search performed as well as SBIDOEMAN BMA in the one-step model, it performs worse in the more complex two-step model, suggesting that principled heuristics and optimal experimental design algorithms are needed for more complex models of biology.

\section{Discussion}

\textbf{Robustness and Performance of SBIDOEMAN BMA} We demonstrated the ability of SBIDOEMAN BMA to select a true model of the BMP pathway over competing methods, including a standard heuristic in biological systems. Across an ensemble of models, prediction of optimal designs and subsequent posterior evaluation are more efficient. In the process of comparing SBIDOEMAN BMA, we have shown how to estimate a model's marginal probability using normalizing flows. We have also shown that averaging the mutual information estimate between models still results in designs that outperform competing methods in improving the quality of experiments. 

\textbf{Future Work} We demonstrated application of SBIDOEMAN BMA to two simple models, the one-step and two-step models, of the BMP pathway, each with two and three parameters, respectively. The efficacy of the SBIDOEMAN algorithm to scale and infer larger implicit likelihoods, such as a BMP model with up to 275 parameters for a BMP two-step model with up to 10 unique ligands, 4 type I receptors, and 3 type II receptors, remains to be seen. Recent innovations in LFI methods, such as SNLVI \cite{Glockler2022}, show promise to ease such problems, demonstrating robust inference of a neuroscience model with 31 parameters. We also only consider the case without noise or batch effects observed in the model, which true biological systems have. Future work will examine how SBIDOEMAN BMA scales to larger models and how robust it is to noise and batch effects in biological systems.

Computationally, we only used averaging of the mutual information among models to design optimal experiments. Weighting each model's mutual information by its respective marginal probability may lead to improved designs for for the model with more evidence. We used a simple ensemble method to evaluate the performance of iid models using our algorithm, which allows us to measure uncertainty in models' predictions. Mixtures of Experts (MoEs) \cite{Bengio2013, Shazeer2017} have also shown promise to improve training and can be combined with ensembling methods to also perform uncertainty quantification \cite{Allingham2021}. These methods could both improve performance and uncertainty quantification in optimal designs for biological models.



\section*{Software and Data}

We used the hydra configuration manager to track hyperparameters and seeds of experiments \cite{Yadan2019Hydra}. To perform SBI, we used the SBI software library \cite{tejero-cantero2020sbi}. The model marginal probability calculation was performed using JAX and Distrax libraries \cite{jax2018github, deepmind2020jax}.

\section*{Acknowledgements}

This research was funded by the National Institute of General Medical Sciences (NIGMS) of the National Institutes of Health (NIH) under award numbers R01GM134418 and 1F31GM145188-01. We would like to thank Matthew Langley for helpful conversations about the BMP pathway, Christina Su for making the simulator that made this work possible, and other members of the Elowitz Lab for helpful discussions. We thank Zelda Mariet for revising our proof.


\bibliography{bibs}
\bibliographystyle{icml2022}

\newpage
\appendix
\onecolumn
\section{Algorithmic implementation}

\vskip -0.1in
\begin{algorithm}[H]
\caption{Bayesian Model Averaging SBIDOEMAN}\label{sdmAlgo}
\begin{algorithmic}[1]
\STATE \textbf{Require:} Simulators $f_i(\boldsymbol{d}, \boldsymbol{\theta})$, held-out true parameters $\boldsymbol{\theta_T}$, true simulator $f_T$,  number of simulations for MINEBED $N_M$, number of acquisitions for Bayesian optimization $N_A$, number of simulations per LFI round $N_S$, number of LFI rounds $N_R$, number of experiments $N_E$, neural density estimator $q_{\phi}(\boldsymbol{x}|\boldsymbol{\theta})$, priors over models' parameters $p(\boldsymbol{\theta}_i)$, and models' prior probabilities $p(\mathcal{M}_i)$.

\STATE \textbf{Return:} Models' approximate posterior $p(\boldsymbol{\theta}_i|\boldsymbol{x_o}, \boldsymbol{d}, \mathcal{M}_i)$, models' marginal probabilities $p(\mathcal{M}_i |\boldsymbol{x_o}, \boldsymbol{\theta}_i, \boldsymbol{d})$, and Bayes Factor $BF = p(\mathcal{M}_1)/p(\mathcal{M}_0)$.

\STATE Initialize a design $\boldsymbol{d_0}$ by random sampling and set $\boldsymbol{d}^*=\boldsymbol{d}_0$ \\
\STATE Initialize $N$ MINE neural network parameters $\boldsymbol{\psi}_0, \dots, \boldsymbol{\psi}_N$ where $N = |\mathcal{M}|$
\STATE Set proposals $\Tilde{p}^{\{i\}}(\boldsymbol{\theta}) \coloneqq p^{\{i\}}(\boldsymbol{\theta})$ for $\mathcal{M}_i \in \{ \mathcal{M}\}^N_{i=1}$

\FOR{$j = 1:N_E$}
    \FOR{$\mathcal{M}_i \in \{ \mathcal{M}\}^N_{i=1}$}
        \FOR{$k = 1:N_A$}
            \FOR{$l = 1:N_M$}
                \STATE{ $\boldsymbol{\theta}_{k,l}^{\{i\}} \sim  \Tilde{p}_{k,l}^{\{i\}}(\boldsymbol{\theta})$ }
                \STATE{ Simulate $\boldsymbol{x}_{k,l}^{\{i\}} \sim f_i(\boldsymbol{d}, \boldsymbol{\theta}_{k,l}^{\{i\}})$}
                \STATE{ Optimize MINE parameters $\psi_i$ between simulated data and priors for the model by maximizing the mutual information lower bound $\widehat{I}(\boldsymbol{d}, \psi_i^*)$ }
                \STATE{ $\widehat{I}(\boldsymbol{d}, \psi^*) = \frac{1}{N}\sum \widehat{I}(\boldsymbol{d}, \psi_i^*)$ }
                \STATE{ $\boldsymbol{d}^* = \boldsymbol{d}$ if $\widehat{I}(\boldsymbol{d}, \psi^*) > \widehat{I}(\boldsymbol{d}^*, \psi^*)$}
            \ENDFOR
        \ENDFOR
    \ENDFOR
    
    \STATE{ Observe simulated experimental condition $\boldsymbol{x_o} = f_T(\boldsymbol{d}^*,\boldsymbol{\theta}_T)$}
    
    \FOR{$\mathcal{M}_i \in \{ \mathcal{M}\}^N_{i=1}$}    
        \FOR{$k = 1:N_R$}
            \FOR{$l = 1:N_S$}
            \STATE{ $\boldsymbol{\theta}_{k,l} \sim \Tilde{p}_k^{\{i\}}(\boldsymbol{\theta})$}
            
            \STATE{ Simulate $\boldsymbol{x}_{k,l} \sim f_i(\boldsymbol{x}, \boldsymbol{\theta}_{k,l})$}
            
            \ENDFOR
        
        \STATE{ (re-)train $q_{\phi}^{\{i\}} \leftarrow \underset{\phi}{\argmin} -\frac{1}{N}\sum_{(\boldsymbol{x}_{k,l}^{\{i\}}, \boldsymbol{\theta}_{k,l}^{\{i\}})}^j \log q_{\phi}^{\{i\}}(\boldsymbol{x}_{k,l}^{\{i\}} |\boldsymbol{\theta}_{k,l}^{\{i\}})$ }
        
        \STATE{  $\Tilde{p}_{k+1}^{\{i\}}(\boldsymbol{\theta}| \boldsymbol{x_o}) \propto p^{\{i\}}(\boldsymbol{x_o}|\boldsymbol{\theta}) p^{\{i\}}_k(\boldsymbol{\theta}) \approx q_{\phi}^{\{i\}}(\boldsymbol{x_o}|\boldsymbol{\theta}) p^{\{i\}}_k(\boldsymbol{\theta})$ }
        
        \STATE{ (re-)train $q_{\tau}^{\{i\}} \leftarrow  \underset{\tau}{\argmin} D_{KL}(q_{\tau}^{\{i\}}(\boldsymbol{\theta})|| \Tilde{p}_{k+1}^{\{i\}}(\boldsymbol{\theta}|\boldsymbol{x_0}))$ }
        
        \STATE{ Set $\Tilde{p}_{k+1}^{\{i\}}(\boldsymbol{\theta}) \coloneqq q_{\tau}^{\{i\}}(\boldsymbol{\theta})$}
        
        \ENDFOR
        
    \STATE{ train $p_x(\boldsymbol{x} | \boldsymbol{\theta}, \boldsymbol{x_o}, \mathcal{M}_i) \leftarrow \underset{\zeta}{\argmin} - \frac{1}{N} \sum_{(\boldsymbol{x}_{l}, \boldsymbol{\theta}_{l})}^j \log p_u(\boldsymbol{f}^{-1}(\boldsymbol{x_o}; \zeta) + \log \rvert \det \boldsymbol{J}(\boldsymbol{f}^{-1})(\boldsymbol{x_o}; \zeta) \rvert$ where $p_u(\boldsymbol{u}) \sim \mathcal{N}(\boldsymbol{0}, \boldsymbol{1})$}
     
    \STATE{ $p(\mathcal{M}_i| \boldsymbol{x_o}, \boldsymbol{\theta}, \boldsymbol{d} ) = 1 - \frac{1}{2}\left(1 + \mathrm{erf}\left(\frac{\boldsymbol{f}^{-1}(\boldsymbol{x_o};\zeta)}{\sqrt{2}}\right)\right)$}
    
    \ENDFOR

\STATE{ $BF_j$ = $\frac{p(\mathcal{M}_1)}{p(\mathcal{M}_0)}$}

\ENDFOR
\end{algorithmic}
\end{algorithm}
\vskip -0.15in
For the choice of hyperparameters, we used $N_M=5000$, $N_A=5$, $N_S=1000$, $N_R = 5$, $N_E = 5$, a SNLE $q_{\phi}(x|\theta)$ density estimator, starting box uniform priors for $p(\theta)$, and uniform priors for $p(\mathcal{M}_i)$.   We evaluated 50 simulations at a time limit of 10 hours. For the one-step model, the random choice had 14 simulations finish, equidistant had 26 simulations finish, and SBIDOEMAN BMA had 15 simulations finish. For the two-step model, random choice had 21 simulations finsih, equidistant had 25 finish, and SBIDOEMAN BMA had 16 finish. 


\section{Proof of Proposition 2.1}

We prove \cref{prop:margProbFlow} using the following definitions. 

\begin{definition}
\label{def:MargLike}
The marginal likelihood is the measure of evidence given a model and is defined as
\begin{equation}
    p(\boldsymbol{x_o}| \mathcal{M} ) = \int p(\boldsymbol{x_o}|\boldsymbol{\theta}, \mathcal{M}) p(\boldsymbol{\theta} | \mathcal{M}) d \boldsymbol{\theta}      \notag
\end{equation}
\end{definition}

where the parameters $\boldsymbol{\theta}$ of the model $\mathcal{M}$ are integrated out. If the marginal likelihood is multiplied by the model prior $p(\mathcal{M})$ it then becomes model marginal probability $p(\mathcal{M}|\boldsymbol{x_o})$.

\begin{definition}
\label{def:NormalizeU}
A normalizing flow is defined as a series of diffeomorphic functions, $\boldsymbol{f}$, that map a base distribution, $p_{\boldsymbol{u}}(\boldsymbol{u})$ to a complex data distribution, $p_{\boldsymbol{x}}(\boldsymbol{x})$. Equation \labelcref{NormFlow} shows the transformation from a base distribution to data distribution. The inverse function $\boldsymbol{f}^{-1}$ will transform observed data $\boldsymbol{x}$ to the base distribution as $\boldsymbol{f}^{-1}(\boldsymbol{x})=\boldsymbol{u}$, which could be a Gaussian distribution. When training a normalizing flow with fixed base distribution parameters, this can also be seen as a generalized version of the reparameterization trick and a form of variational inference \cite{Rezende2015}.
\end{definition}

\begin{definition}
\label{def:GaussianCDF}
The Cumulative Distribution Function (CDF) of a real-valued random variable $X$, if $X$ is distributed according to a Gaussian distribution, is defined as 
\begin{align*}
    F_X(x) &= P(X \leq x) \\
    &= \Phi(x) \\
    &= \int_{-\infty}^x \frac{1}{\sqrt{2 \pi}} \exp{\left(-\frac{1}{2} u^2\right)} du \\
    &= \frac{1}{2}\left(1 + \mathrm{erf}\left(\frac{x}{\sqrt{2}}\right)\right)
\end{align*}

where $\mathrm{erf}()$ is the error function. A derivation of the Gaussian CDF can be found in \citet{bishop2006pattern}.
\end{definition}

\begin{definition}
\label{def:ProbSurvival}
The CDF gives a probability of points smaller than a point but we want to know how many points are larger than the point in the Gaussian distribution, $P(X > x) = 1 - F_X(x)$. This is the complementary CDF, or tail distribution, and is helpful in the context of normalizing flows, where we are interested in determining the probability of simulated point from our model $\boldsymbol{x} \sim p_x(\boldsymbol{x}| \boldsymbol{x_o}, \boldsymbol{\theta}, \mathcal{M})$ being greater than the observed data point $\boldsymbol{x_o}$. Thus, $p( \boldsymbol{x_o} | \mathcal{M}, \boldsymbol{\theta}) \equiv 1 - F_X(\boldsymbol{x_o})$.
\end{definition}

Using these definitions, we prove \cref{prop:margProbFlow} as follows.

\begin{proof} 
Starting from the definition of the marginal likelihood in \cref{def:MargLike}, we can approximate the intractable likelihood $p(\boldsymbol{x}|\boldsymbol{\theta}, \mathcal{M})$ using a normalizing flow trained by sampling $\boldsymbol{u} \sim \mathcal{N}(\boldsymbol{0}, \boldsymbol{1})$, as by variational inference in \cref{def:NormalizeU}, and model parameters sampled $\theta \sim q_{\tau}(\theta)$ where $q_{\tau}(\theta)$ is the model's posterior given observed data $x_o$. Thus the marginal likelkihood can be approximated as follows:
\begin{align*}
    p(\boldsymbol{x}| \mathcal{M} ) &= \int p(\boldsymbol{x}|\boldsymbol{\theta}, \mathcal{M}) p(\boldsymbol{\theta}, \mathcal{M}) d \boldsymbol{\theta} && \text{By \cref{def:MargLike}}    \\
    &\approx p_x(\boldsymbol{x} | \boldsymbol{\theta}, \mathcal{M}) && \text{By \cref{def:NormalizeU} and } \boldsymbol{\theta} \sim  q_{\tau}(\boldsymbol{\theta} | \boldsymbol{x_o},  \mathcal{M}).
\end{align*}
The last step can also be seen as a form of variational inference that employs the reparameterization trick for a simulator $f_{\mathcal{M}}$ as $\boldsymbol{x} \sim p_x(\boldsymbol{x} | \boldsymbol{\theta}, \mathcal{M}, \boldsymbol{x_o}) \Longleftrightarrow \boldsymbol{x} = f_{\mathcal{M}}(\boldsymbol{d}, \boldsymbol{\theta}), \boldsymbol{\theta} \sim q_{\tau}(\boldsymbol{\theta} | \boldsymbol{x_o}, \mathcal{M})$, assuming that designs $\boldsymbol{d}$ are known and using the posterior inference of model parameters given a certain observed data point $\boldsymbol{x_o}$. 

Using this approximation of the marginal probability distribution, we can find the probability that a point $\boldsymbol{x_o}$ is within this probability density function by converting to a Gaussian base distribution, $p_u(\boldsymbol{u})$, via variational inference. We find models' marginal probabilities as follows: 
\begin{align*}
    p( \boldsymbol{x_o} | \mathcal{M}, \boldsymbol{\theta} ) &\equiv 1 - F_X(\boldsymbol{x_o}) && \text{By \cref{def:ProbSurvival}} \\
    &= 1 - \int_{-\infty}^{\boldsymbol{x_o}} p_x(\boldsymbol{x} |\boldsymbol{\theta}, \mathcal{M}, \boldsymbol{d} )d\boldsymbol{x} \\
    &= 1 - \int_{-\infty}^{\boldsymbol{x_o}} p_u(\boldsymbol{f}^{-1}(\boldsymbol{x})) \rvert \det \boldsymbol{J}(\boldsymbol{f}^{-1})(\boldsymbol{x}) \rvert   d\boldsymbol{u} && \text{By \cref{def:NormalizeU}} \\
    &\propto 1 - \int_{-\infty}^{\boldsymbol{x_o}} p_u(\boldsymbol{f}^{-1}(\boldsymbol{x})) d\boldsymbol{u} \\
    &= 1 - \Phi(\boldsymbol{f}^{-1}(\boldsymbol{x_o})) && \text{By \cref{def:GaussianCDF}}\\
    &= 1 - \frac{1}{2}\left(1 + \mathrm{erf}\left(\frac{\boldsymbol{f}^{-1}(\boldsymbol{x_o})}{\sqrt{2}}\right)\right) && \text{By \cref{def:GaussianCDF}}.
\end{align*}

We note that this data likelihood becomes the model marginal probability when multiplied by the model's prior, $p(\mathcal{M})$.
\end{proof}

\begin{remark}
While we achieved acceptable results by simplifying the change of volume from the data distribution $p_x(x)$ to the base distribution $p_u(u)$ in the third step, it remains to be seen whether there is improved performance in calculating models' marginal probabilities with this value.
\end{remark}

\end{document}